\documentclass[aps,prl,nopacs, twocolumn, nofootinbib]{revtex4-1}   

\usepackage[utf8]{inputenc}  
\usepackage[british]{babel}  
\usepackage{palatino}
\usepackage[colorlinks,citecolor=blue,linkcolor=blue,urlcolor=blue]{hyperref}  
\usepackage{graphicx} 
\usepackage{amsmath,amssymb,amsthm,bm,mathtools,amsfonts,mathrsfs,bbm,dsfont} 

\newcommand{\bra}[1]{\langle #1|}
\newcommand{\ket}[1]{|#1\rangle}
\newcommand{\braket}[2]{\langle #1|#2\rangle}
\newcommand{\ketbra}[1]{| #1\rangle \langle #1|}

\newcommand{\be}{\begin{equation}}
\newcommand{\ee}{\end{equation}}
\newcommand{\bea}{\begin{eqnarray}}
\newcommand{\eea}{\end{eqnarray}}

\newcommand{\mean}[1]{\ensuremath{\langle{#1}\rangle}}

\newcommand{\eins}{\mathbbm{1}}
\newcommand{\WW}{\ensuremath{\mathcal{W}}}

\newcommand{\SSS}{\ensuremath{\mathcal{S}}}

\newcommand{\kommentar}[1]{}
\newcommand{\tr}{{\rm Tr}}

\renewcommand{\vr}{\ensuremath{\varrho}}

\newcommand{\forget}[1]{}

\begin{document}

\title{Geometry of faithful entanglement}

\author{Otfried Gühne}
\email{otfried.guehne@uni-siegen.de}
\affiliation{Naturwissenschaftlich-Technische Fakultät, 
Universität Siegen, Walter-Flex-Straße 3, 57068 Siegen, Germany}

\author{Yuanyuan Mao}
\email{yuanyuan.mao@uni-siegen.de}
\affiliation{Naturwissenschaftlich-Technische Fakultät, 
Universität Siegen, Walter-Flex-Straße 3, 57068 Siegen, Germany}

\author{Xiao-Dong Yu}
\email{xiao-dong.yu@uni-siegen.de}
\affiliation{Naturwissenschaftlich-Technische Fakultät, 
Universität Siegen, Walter-Flex-Straße 3, 57068 Siegen, Germany}

\date{\today}

\begin{abstract}
A typical concept in quantum state analysis is based on 
the idea that states in the vicinity of some pure entangled 
state share the same properties; implying that states with 
a high fidelity must be entangled. States whose entanglement 
can be detected in this way are also called faithful. We prove 
a structural result on the corresponding fidelity-based entanglement witnesses, resulting in a simple condition for faithfulness of a two-party state. For the simplest case of two qubits faithfulness can directly
be decided and for higher dimensions accurate analytical criteria
are given. Finally, our results show that faithful entanglement 
is, in a certain sense, useful entanglement; moreover, they establish
connections to computational complexity and simplify several results 
in entanglement theory.
\end{abstract}
\maketitle

{\it Introduction.---}
The certification and benchmarking of quantum devices is 
a fundamental problem in quantum information processing 
\cite{klieschprxq, eisertnatrev}. Here, the exponentially
increasing dimensionality of large quantum systems makes 
a full characterization usually impossible. Hence, one 
often restricts the attention to simple parameters, such 
as the average gate fidelity of a quantum gate 
\cite{schumacher96, nielsengate}, in order to compare different 
approaches or implementations. For the justification of 
such restrictions it is necessary, however, to know 
which phenomena are captured by the considered parameter 
and which not.

For the certification of quantum states, the fidelity of
the observed state with a desired target state is often
used as a benchmarking parameter. This can indeed be 
measured or estimated with relatively simple methods 
\cite{flammia, guehfidest}. A natural question is then 
how this parameter is connected to the entanglement in 
the quantum state under scrutiny. This is relevant since
entanglement is a key ingredient for applications like
quantum key distribution \cite{curtylutkenhaus} or 
quantum metrology \cite{pezzesmerzi}. 

A typical method to connect entanglement to the fidelity 
is based on the idea that states close to some known 
entangled pure state $\ket{\psi}$ must be entangled, too. 
Formally, the fidelity can be seen as a kind of distance 
of a general state $\varrho$ to the state $\ket{\psi}$, 
given by $F_\psi = \bra{\psi}\varrho \ket{\psi}.$ Then, 
one can formulate the resulting criteria also in the form 
of entanglement witnesses. That is, one considers the 
observable
\be
\WW = \alpha \eins - \ketbra{\psi}.
\label{eq-intro}
\ee
In general, an entanglement witness is an observable which
has a positive mean value on all separable states, hence
a negative mean value signals the presence of entanglement
\cite{hororeview, gtreview, friisreview}. 
Concretely, if this observable $\WW$ is measured, one obtains 
$\tr(\varrho \WW)= \alpha - F_\psi$, so if $F_\psi$ is above
the threshold value $\alpha$, then the witness detects some
entanglement. The fidelity-based witness in Eq.~(\ref{eq-intro})
is easy to construct and can be generalized to the 
multiparticle case. Note, however, that many other approaches 
to construct witnesses beyond the fidelity approach exist
\cite{terhal, optimization, gezaspin, lurccnr, pianistates, 
dariusreview}.

So, it is a central question to ask which quantum states can be 
detected as entangled using the fidelity and what are their 
{\it physical} properties. In Ref.~\cite{weilenmann20}, the 
authors coined the term {\it faithful} for states whose entanglement 
can be characterized using the fidelity, and provided first steps in distinguishing faithful and unfaithful states. For instance, an approach using convex optimization was provided which can prove
the unfaithfulness of a state, with this one can show that in higher 
dimensional systems most states are unfaithful. Very recently, the
phenomenon of faithfulness was also studied from an experimental 
point of view \cite{hu2020, riccardi2021}.

In this paper we go further and deliver analytical results on faithful
and unfaithful states. First, we show that one can restrict the attention
to a special subclass of fidelity-based criteria. Then, we can derive some
general results on faithful states: For two qubits, the faithfulness can
directly be decided, and for higher-dimensional systems, an analytical 
sufficient criterion for unfaithfulness can be given. We also show that faithful entanglement is useful entanglement, in the sense that it leads 
to a violation of Bell inequalities on multiple copies of a state.
Moreover, our results simplify estimates of entanglement measures 
and we demonstrate that entanglement detection using the fidelity of 
a general projector $\Pi$ instead of a pure state $\ketbra{\psi}$  is fundamentally different. Finally, we establish a connection to the 
problem of unitary quadratic minimization in complexity theory, as 
well as possible extensions to the Schmidt number classification of quantum states.

{\it Entanglement, separability and witnesses.---}
We consider two parties, Alice and Bob, each of which owns 
a $d$-dimensional quantum system. In general, a mixed state 
is called separable, if it can be  written as a convex 
combination of product states, 
$
\varrho_{AB} = \sum_k p_k \ketbra{a_k, b_k},
$
where the $p_k$ form a probability distribution, that is, they 
are positive and
sum up to one. States that can not be expressed in this way 
are called entangled. 

There are many criteria for proving entanglement or separability of a given
quantum state, although none of them solves the problem completely 
\cite{hororeview, gtreview}. An important criterion is entanglement 
witnesses, which are observables that have a positive expectation 
value on all separable states, and a negative expectation value on 
some entangled states. It can be straightforwardly shown that, 
in principle, any entangled state can be detected by some suitable 
witness, but the problem is to construct all witnesses. This is, in 
fact, not easy and of the same complexity as the separability problem.

As already mentioned, one possible witness construction makes use of the 
fact that in the vicinity of an entangled pure state there are only entangled 
states. So, one can define a witness via $\WW = \alpha \eins - \ketbra{\psi},$
where $\alpha$ is the maximal squared overlap between $\ket{\psi}$ and the 
separable states. This can be computed as 
$
\alpha = 
\sup_{\ket{a,b}} |\braket{a,b}{\psi}|^2
=s_1^2,
$
where the $s_i$ for $i = 1, \dots, r$ are the decreasingly ordered 
nonzero Schmidt coefficients of the state $\ket{\psi}$ in its Schmidt 
decomposition $\ket{\psi}=\sum_{i=1}^r s_i \ket{ii}$ \cite{bourennane}. 
Here, the number of terms $r$ is called the Schmidt rank of the state. 
The smallest possible $\alpha$ occurs if the state $\ket{\psi}$
is a maximally entangled state, for instance
$\ket{\psi}=\ket{\phi^+}=\sum_{i=1}^d \ket{ii}/\sqrt{d}.$ 
Then we have 
\be
\WW = \frac{\eins}{d} - \ketbra{\phi^+}.
\label{eq-rfw}
\ee 
Clearly, one may also take other maximally entangled states which 
are locally equivalent to $\ket{\phi^+}$. For reasons that 
become apparent later, we call this set of witnesses of the 
type in Eq.~(\ref{eq-rfw}) the relevant fidelity witnesses (RFW).

{\it Main result.---}
We can directly state and prove the first main result:

{\bf Observation 1.} {\it Let $\varrho_{AB}$ be a faithful entangled 
state, i.e. its entanglement can be detected by some fidelity-based 
entanglement witness. Then $\varrho_{AB}$ can be detected by a relevant 
fidelity-based entanglement witness. In other words, a state $\varrho_{AB}$ 
is faithful if and only if there are local unitary transformations $U_A$ 
and $U_B$ such that
\be
\bra{\phi^+}U_A \otimes U_B\varrho_{AB}U_A^\dagger \otimes U_B^\dagger\ket{\phi^+} > \frac{1}{d}.
\label{formula}
\ee
}

In order to present the idea of the proof we consider the case 
$d=4$, the general case is discussed in Appendix A. 
Let us assume that the fidelity-based witness detecting 
$\varrho_{AB}$ is given by
$
\WW = s_1^2 \eins - \ketbra{\psi},
$
with $\ket{\psi}= \sum_{i=1}^4 s_i \ket{ii}.$ We consider eight RFWs, 
coming from the eight maximally entangled states
\be
\ket{\phi_{\vec{a}}} = \frac{1}{2}
\big(
\ket{11}+ a_2 \ket{22}+ a_3 \ket{33}+ a_4 \ket{44}
\big),
\ee
where $\vec{a}=(a_2,a_3,a_4)$ and the coefficients $a_j$ can 
have all possible values $\pm 1$. This leads to eight RFWs
$\WW_{\vec{a}} = \eins/4 - \ketbra{\phi_{\vec{a}}}$.

We aim to show that one can find probabilities $p_{\vec{a}}$ such that
the operator
\be
\mathcal{Z}=\WW - 4 s_1^2 \sum_{{\vec{a}}}  p_{\vec{a}} \WW_{\vec{a}} \geq 0
\label{eq-proof1}
\ee
is positive semidefinite. This would prove already the main claim, 
since then $\tr(\varrho_{AB} \WW) < 0$ implies that for at least 
one ${\vec{a}}$ we have $\tr(\varrho_{AB} \WW_{\vec{a}}) < 0$.
In order to prove Eq.~(\ref{eq-proof1}) we aim to find $p_{\vec{a}}$ such that
$\mathcal{Z}$ is diagonal. This is sufficient for positivity, as the diagonal
entries of $\mathcal{Z}$ are independent of the $p_{\vec{a}}$. They are either 
zero
or of the type $s_1^2-s_j^2 \geq 0,$ which ensures $\mathcal{Z} \geq 0.$

Looking at the relevant off-diagonal entries, one finds that 
for making $\mathcal{Z}$ diagonal, we have to find $p_{\vec{a}}$ such that
\begin{widetext}
\be
\begin{pmatrix}
\times & \alpha_2 & \alpha_3 & \alpha_4 \\
\alpha_2 & \times & \alpha_2 \alpha_3 & \alpha_2 \alpha_4\\
\alpha_3 &  \alpha_2 \alpha_3 &  \times &\alpha_3\alpha_4 \\
\alpha_4 & \alpha_2 \alpha_4&\alpha_3\alpha_4&  \times
\end{pmatrix}
= \sum_{{\vec{a}}} p_{\vec{a}}
\begin{pmatrix}
\times & a_2 & a_3 & a_4 \\
a_2 & \times &a_2 a_3 & a_2 a_4\\
a_3 &  a_2 a_3 &  \times &a_3 a_4 \\
a_4 & a_2 a_4 &a_3 a_4&  \times
\end{pmatrix},
\label{eq-offdiagonal}
\ee
\end{widetext}
where we defined
$
\alpha_i = {s_i}/{s_1} \in [0,1].
$
With a physical argument one can see now that a decomposition as in 
Eq.~(\ref{eq-offdiagonal}) can be found: One can view the three 
$\alpha_i$ as expectation values of some observables (say, $\sigma_z$)
on a three-qubit product state $\vr=\vr_A\otimes\vr_B\otimes\vr_C$. The 
terms $\alpha_i \alpha_j$ correspond then to a two-body correlation 
$\mean{\sigma_z \otimes \sigma_z}$ on the same state.  On the other 
hand, the right-hand side of Eq.~(\ref{eq-offdiagonal}) can be seen 
as a local hidden variable model, where the index ${\vec{a}}$ is the hidden 
variable occurring with probability $p_{\vec{a}}$, and the $a_i$ are 
the deterministic assignments for the measurement results of $\sigma_z$ 
on the different particles. For fully separable states, however, 
it is well known that all measurements can be explained by a local 
hidden variable model \cite{werner}, so there must be $p_{\vec{a}}$ such that 
Eq.~(\ref{eq-offdiagonal}) is fulfilled.
\qed

{\it General implications.---}
First, from Observation 1 it is immediately clear that already for
two qubits not all entangled states are faithful. The reason is 
the following: The RFW in Eq.~(\ref{eq-rfw}) can be seen as a witnesses 
for the computable cross-norm or realignment (CCNR) criterion 
\cite{chen, rudolph}. This follows from the fact that one may write 
it as  
\be
\WW = \frac{1}{d}
\big(
\eins - \sum_{k=1}^{d^2} G_k \otimes G_k^T
\big),
\ee
where the $\{G_k\}$ form an orthogonal basis of the operator 
spaces for Alice and Bob and $(\cdot)^T$ denotes
the transposition \cite{lurccnr}. This means that
only states that violate the CCNR criterion can be faithful.
But it is known that already for two qubits there are entangled
states that do not violate this criterion \cite{rudolphpra}.

Second, using the Jamio{\l}kowski isomorphism \cite{depillis, jamiol}, 
one can directly see that the RFW in Eq.~(\ref{eq-rfw}) corresponds
to the reduction map. From this one has directly the already known
result \cite{pianistates} that only the entanglement in states which 
violate the reduction criterion can be detected by the fidelity.
This further implies that only states that violate the criterion of 
the positivity of the partial transpose (PPT) can be faithful.

Third, the left hand side of the inequality (\ref{formula}) is related 
to a well-known quantity called the singlet fraction. This concept was 
introduced in Ref.~\cite{singfrac} in a slightly more general form, 
where one tries to maximize the fidelity with the maximally entangled state, using trace-preserving local quantum operations and classical communication. Then, it was proved that having a singlet fraction larger than $1/d$  is a necessary and sufficient condition for a state to offer a quantum advantage in teleportation. 
It  was also  shown  that  for  every  state $\varrho$ with singlet fraction larger than $1/d$, there exists a number of copies $k$ of $\varrho$ such that 
$\varrho^{\otimes k}$ is nonlocal \cite{sfnl}. Together with these results, 
Observation 1 implies that faithful states are useful for quantum teleportation
and that multiple copies of them violate a Bell inequality. This shows that 
fidelity-based  entanglement witnesses detect entanglement that is useful.

Finally, the connection to the singlet fraction delivers more insights: For 
two qubits this quantity can be lower bounded by the entanglement as 
quantified by the concurrence or negativity \cite{verstraetefidelity}. 
From these results one can directly conclude that if the concurrence 
exceeds $1/2$ or the negativity exceeds $(\sqrt{2}-1)/2$, then the state 
is faithful. Also, Ref.~\cite{verstraete} implies that any two-qubit 
state is faithful after suitable local filtering operations.
Moreover, as we elaborate in Appendix B, 
the unitary optimization in Eq.~(\ref{formula}) is directly linked to 
the problem of unitary quadratic minimization, which was recently
shown to be NP-hard \cite{lee2020}.

{\it Deciding faithfulness for qubits.---}
The problem of maximizing the overlap with a maximally entangled 
state in some basis for two qubits has been solved before 
\cite{badziag}, but in order to generalize it later, 
we formulate it in a different language. We can decompose 
a general two-qubit state $\varrho_{AB}$ into Pauli matrices, 
\be
\varrho_{AB} = \frac{1}{4} \sum_{i,j = 0}^{3} \lambda_{ij} \sigma_{i} \otimes \sigma_j,
\label{eq-bloch}
\ee
where $\lambda_{ij} = \tr(\varrho \sigma_{i} \otimes \sigma_j)$ 
and $\sigma_0= \eins$ denotes the identity matrix. Then we consider 
the operator
\be
X_d(\varrho_{AB}) = \varrho_{AB} -\frac{1}{d}(\varrho_A \otimes \eins + \eins \otimes \varrho_B) 
+ \frac{2}{d^2}\eins\otimes \eins,
\label{eq-xdef}
\ee
where, for the case of two qubits, we take $d=2.$ In terms of the representation in 
Eq.~(\ref{eq-bloch}) this operator has a block-diagonal $\lambda_{ij}$, the terms 
$\lambda_{i0}$ and $\lambda_{0j}$ for $i,j = 1,2,3$  corresponding 
to the marginals have been removed. Note that we have for any maximally 
entangled state in any basis 
$\bra{\phi^+} X_2(\varrho)\ket{\phi^+} = \bra{\phi^+} \varrho_{AB} \ket{\phi^+}$, 
as for maximally entangled states the marginals are maximally 
mixed. 

By local unitary transformations we can diagonalize the 
remaining $3 \times 3$ matrix $\lambda_{ij}$ for $i,j = 1,2,3$ 
for the operator $X_2$. In this basis, $X_2$ is Bell diagonal, 
and we can directly read of the maximal overlap with Bell states 
by computing the maximal eigenvalue. We can summarize: 

{\bf Observation 2.} 
{\it
A two-qubit state $\varrho_{AB}$ is faithful if and only if the maximal 
eigenvalue of $X_2(\varrho_{AB})$ in Eq.~(\ref{eq-xdef}) is larger than 
$1/2$. 
}

{\it Deciding faithfulness for higher dimensions.---}
First, we consider $X_d(\varrho_{AB})$ from above, 
and if its largest eigenvalue is smaller than $1/d$, the state
cannot be faithful. The proof of this statement follows along 
the same lines as for qubits, one just needs to replace the Pauli 
matrices by some other basis of the operator space, where one basis 
element is proportional to the identity and the other elements
are traceless. Then, the block of the matrix $\lambda$ can, in
general, not be diagonalized anymore. Still, the maximal eigenvalue 
of $X_d$ is an upper bound on the overlap with maximally entangled 
states.

Alternatively, instead of maximizing the overlap over all maximally 
entangled states, one can maximize the overlap over all states with 
maximally mixed reduced states. That is, one considers the relaxed 
optimization
\begin{align}
\mbox{max: } & \tr(\varrho_{AB} \chi),
\label{eq-maxddim}
\\
\mbox{subject to: } & \tr(\chi) = 1, \quad \chi \geq 0,
\\
& \tr_A(\chi) = \tr_B (\chi) = \frac{\eins}{d}.
\end{align}
This is a simple SDP, and if a result smaller than $1/d$ is found, the
state cannot be faithful. In fact, one can show that the dual of this 
SDP is equivalent to the SDP2 presented in Ref.~\cite{weilenmann20}. 

The formulation in  Eq.~(\ref{eq-maxddim}) has, however, two advantages.
First it can also be used to prove that a state is faithful. If the SDP 
returns a value larger than $1/d$ one can check whether the optimal 
$\chi$ is a pure state. If this is the case, this state is maximally 
entangled  and $\varrho_{AB}$ must be faithful. Second, one can 
systematically improve the SDP in Eq.~(\ref{eq-maxddim}) by adding 
rank constraints on $\chi.$ These can be implemented by a hierarchy of 
SDPs \cite{xiaodongnew} and detect indeed more states; see also below.
We summarize:

{\bf Observation 3.} 
{\it
Consider a general two-qudit state $\varrho_{AB}$.
(a) If the largest eigenvalue of the operator $X_d(\varrho_{AB})$ 
in Eq.~(\ref{eq-xdef}) is not larger than $1/d$, then 
$\varrho_{AB}$ is unfaithful.
(b) If the semidefinite program from Eq.~(\ref{eq-maxddim}) has 
an optimal value not larger than $1/d$, then $\varrho_{AB}$
is unfaithful. 
(c) If the optimization in Eq.~(\ref{eq-maxddim}) returns a value 
larger than $1/d$ and the optimal $\chi$ is a pure state, then 
$\varrho_{AB}$ is faithful.
}

We add that the condition 3(b) detects strictly more states
as unfaithful than condition 3(a). The reason is that in 
Eq.~(\ref{eq-maxddim}) one can directly replace $\varrho_{AB}$ 
with $X_d$ from Eq.~(\ref{eq-xdef}), without changing the result 
of the SDP. Then, the optimization can further be relaxed by 
considering only the largest eigenvalue of $X_d$. 


{\it Numerical studies.---}
Armed with these insights, we generated random states 
in the Bures metric to estimate the fraction of
the unfaithful states \cite{karolz}, additional results
for the Hilbert-Schmidt distribution are described in
Appendix C. For two qubits, 
we generated $10^6$ states and found $7.32\%$ of all 
states separable via the condition of the positivity
of the partial transpose (PPT) \cite{hororeview} and  
$15.44\%$ entangled, but unfaithful. The remaining 
$77.24\%$ are faithful. These values coincide up to 
statistical fluctuations with the values reported 
for the sufficient criterion SDP2 for unfaithfulness 
in Ref.~\cite{weilenmann20} (or Eq.~(\ref{eq-maxddim})). 
Indeed, using the fact that for $d=2$ the extreme points 
of the unital maps are unitary maps \cite{landau}, one 
can see that in this case the extreme points of the 
matrices $\chi$ considered in  Eq.~(\ref{eq-maxddim}) 
are the projectors onto maximally entangled states, 
hence for $d=2$ the SDP2 in Ref.~\cite{weilenmann20} 
or Observation 3(b) are necessary and sufficient for 
faithfulness.

For higher dimensions the results are given in Table I. In 
the underlying samples of states it turned out that any 
state not obeying the criterion in Observation 3(b) is 
faithful. This, however, is not generally true. Consider 
a highly entangled state $\varrho_{AB}$, where the relaxed 
optimization in Eq.~(\ref{eq-maxddim}) gives a strictly 
larger value than the optimization of the overlap with 
all maximally entangled state, see Eq.~(\ref{formula}), 
but both values are larger than $1/d$. Such states can 
be found by random search. Then, mixing the state with 
white noise leads to a linear decrease of the results in 
both optimizations. For a proper amount of noise, the 
value of Eq.~(\ref{formula}) will be smaller than $1/d$ 
and the value of Eq.~(\ref{eq-maxddim}) strictly larger 
than $1/d$. So, the state will be unfaithful, but the 
criterion of Observation 3(b) does not detect it. In fact, 
in an additional numerical analysis, we also identified
one state for $d=4$ which escapes the detection as 
unfaithful via the criterion 3(b). This state could then 
be detected as unfaithful by the SDP in Eq.~(\ref{eq-maxddim}) 
with additional rank constraints on $\chi$ \cite{xiaodongnew}.

\begin{table}[t]
\begin{tabular}{|c||c|c|c|c|c|c|c|}
\hline
d & PPT & UFF [3(a)] &  UFF [3(b)] & FF [3(c)] &FF [Eq.~(\ref{formula})]\\
\hline
3 & 0\% &25.79\% & 54.68\% & 45.32\% & --- \\
\hline
4 &0\% &71.40 \% &96.959\% & 3.04\% & 0.001\%\\
\hline
5 &0\% & 99.33\% &99.998\% &0.002\% &--- \\
\hline
6 &0\% & 100\% &100\% &0\% &---  \\
\hline \hline
\end{tabular}
\caption{Fraction of states in the Bures metric
that can be detected with the various criteria, 
based on a sample of $10^6$ states for each $d$.
First, we consider the PPT states, which are always 
unfaithful. Then, we consider the NPT states that 
are unfaithful (UFF) due to Observation 3(a). Then, 
the NPT states that are unfaithful due to Observation 
3(b). The fourth column is the fraction of states that 
can be detected as faithful via Observation 3(c). For 
$d=4$ it happened that few states were left where 
faithfulness could not be decided with the previous 
criteria. All of these states could be shown to be
faithful via a direct optimization in Eq.~(\ref{formula}) 
(fifth column).}
\end{table}

{\it Further implications.---}
Among the numerous works on entanglement estimation, many approaches 
use the comparison with the fidelity of some pure entangled states.  
For example, in Ref.~\cite{zhang20} several lower bounds of 
entanglement measures given, which were all of
the type $E(\varrho) \geq f[S(\varrho_{AB})-1]$, 
where $E(\cdot)$ is some entanglement measure, $f[\cdot]$ 
is some function, and
\begin{equation}
S(\varrho_{AB}) = 
\max_{\ket{\psi}}
(\bra{\psi}\varrho_{AB} \ket{\psi}/s_1^2,1),
\end{equation}
which demands the maximization over all pure states. Clearly, $S>1$, 
if and only if $\varrho$ is faithful. From Eq.~(\ref{eq-proof1}) it follows
that for computing $S$ one needs only to optimize over maximally
entangled states. Furthermore, we can conclude that $S>1$ can only
happen if a state violates the CCNR criterion, and Observation 3
can also be applied to estimate $S$.

Also, note that pure states are rank-one projectors and the fidelity 
of a bipartite state $\varrho_{AB}$ and a pure state $\ket{\psi}$ is 
nothing but the expectation value $\bra{\psi}\varrho_{AB} \ket{\psi}$. 
One can thus generalize the definition of the witness in 
Eq.~(\ref{eq-intro}) to
\be
\WW_V= \varepsilon \eins - \Pi_V,
\label{eq-def2}
\ee
where $\Pi_V$ is the projector to some subspace $V$. The minimal value of 
$\varepsilon$ while $\WW$ having positive expectation value on all separable 
states is given by
$\varepsilon_{\rm min}=\sup_{\ket{\psi} \in V, \varrho_{\rm sep}} \bra{\psi} \varrho_{\rm sep} \ket{\psi}$,
which is the maximal value of the largest Schmidt coefficient of the pure states 
in $V$ \cite{sarbicki}. With this construction, also entanglement measures
can be estimated \cite{russian20}. It should be noted, however, that this generalization allows the detection of some bound entangled states with
positive partial transpose, e.g., if $V$ is the subspace complementary to an unextendible product basis \cite{terhal, sarbicki}. In contrast, as we have seen,
the witness in Eq. (\ref{eq-intro}) can only detect entangled states with negative partial transpose. So, the witnesses in Eq.~(\ref{eq-def2}) are structurally
quite different from fidelity-based entanglement witnesses. 

Finally, one may try to extend the notion of relevant fidelity-based witnesses
to the characterization of the Schmidt number \cite{sanpera, hulpke}, this is 
discussed in Appendix~D.

{\it Conclusion.---}
The question whether the entanglement of a state can be characterized 
by the fidelity with some pure state is practically relevant and is 
closely related to the geometry of entangled states. We have provided 
a structural result on such faithful states for bipartite systems.
For the case of two qubits, we provided a simple analytical criterion
for faithfulness, and for higher dimensional systems strong necessary 
criteria and sufficient criteria were developed. Our results showed that fidelity-based
entanglement witnesses detect a form of useful entanglement, moreover, 
they shed light on several other results in entanglement theory, such 
as the estimation of entanglement measures. 

For further work, it would be very interesting to generalize
the approach discussed here to multiparticle systems, where it 
may find applications in the certification of quantum simulation 
\cite{qpt1, qpt2}. Also, one 
may consider other quantum resources (such as measurements or quantum
channels) and discuss the question, which of their properties can be 
inferred by comparing it with a desired ``pure'' resource (e.g., 
a projective measurement or a unitary channel). This may lead 
to new insights into the geometry of these quantum resources.

We thank M. Navascu\'es and M. Weilenmann for discussions. 
This work was supported by the
Deutsche Forschungsgemeinschaft (DFG, German Research Foundation, Project 
No.~447948357 and No.~440958198), the Sino-German Center for Research Promotion, 
and the ERC (Consolidator Grant 683107/TempoQ). Y. M. acknowledges funding from a 
CSC-DAAD scholarship.

%
\section*{Appendix A: Detailed proof of Observation 1}
In this appendix we present the detailed proof of Observation 1
for a general dimension $d$. Let us assume that the fidelity-based 
witness detecting 
$\varrho_{AB}$ is given by
$
\WW = s_1^2 \eins - \ketbra{\psi},
$
with $\ket{\psi}= \sum_{i=1}^d s_i \ket{ii}.$ We consider $2^{d-1}$ 
RFWs, coming from the maximally entangled states
\be
\ket{\phi_{\vec{a}}} = \frac{1}{\sqrt{d}}
\big(
\ket{11}+ \sum_{j=2}^d a_j \ket{jj}
\big),
\ee
where $\vec{a}=(a_2,a_3,\dots, a_d)$ and the coefficients $a_j$ can 
have all possible values $\pm 1$. This leads to $2^{(d-1)}$ RFWs
$\WW_{\vec{a}} = \eins/d - \ketbra{\phi_{\vec{a}}}$.

We aim to show that one can find probabilities $p_{\vec{a}}$ such that
the operator
\be
\mathcal{Z}=\WW - d s_1^2 \sum_{{\vec{a}}} p_{\vec{a}} \WW_{\vec{a}} \geq 0
\label{eq-proof1-appendix}
\ee
is positive semidefinite. This would prove already the main claim, 
since then $\tr(\varrho_{AB} \WW) < 0$ implies that for at least 
one ${\vec{a}}$ we have $\tr(\varrho_{AB} \WW_{\vec{a}}) < 0$.
In order to prove Eq.~(\ref{eq-proof1-appendix}) we aim to find $p_{\vec{a}}$ such that
$\mathcal{Z}$ is diagonal. This is sufficient for positivity, as the 
diagonal entries of $\mathcal{Z}$ are independent of the $p_{\vec{a}}$. They 
are either zero or of the type $s_1^2-s_j^2 \geq 0,$ which ensures 
$\mathcal{Z} \geq 0.$

Looking at the relevant off-diagonal entries, one finds that 
for making $\mathcal{Z}$ diagonal, we have to find $p_{\vec{a}}$ such that
\begin{widetext}
\be
\begin{pmatrix}
\times & \alpha_2 & \alpha_3 & \dots &\alpha_d \\
\alpha_2 & \times & \alpha_2 \alpha_3 & \dots & \alpha_2 \alpha_d\\
\alpha_3 &  \alpha_2 \alpha_3 &  \times & \dots  &\alpha_3\alpha_d \\
\vdots & \vdots & \vdots & \ddots & \vdots\\
\alpha_d & \alpha_2 \alpha_d&\alpha_3\alpha_d& \dots & \times
\end{pmatrix}
= \sum_{{\vec{a}}} p_{\vec{a}}
\begin{pmatrix}
\times & a_2 & a_3 & \dots & a_d \\
a_2 & \times &a_2 a_3 & \dots & a_2 a_d\\
a_3 &  a_2 a_3 &  \times &\dots & a_3 a_d \\
\vdots & \vdots & \vdots & \ddots & \vdots\\
a_d & a_2 a_d&a_3a_d&\dots &  \times
\end{pmatrix},
\label{eq-offdiagonal-appendix}
\ee
\end{widetext}
where we defined
$
\alpha_i = {s_i}/{s_1} \in [0,1]
$
for $i = 2, \dots, d.$

With a physical argument one can see now that a decomposition as in 
Eq.~(\ref{eq-offdiagonal-appendix}) can be found: One can view the  
$\alpha_i$ as expectation values of some observables (say, $\sigma_z$)
on a product state on $d-1$ qubits, i.e., a state of the form 
$\vr=\vr_1\otimes\vr_2\otimes\dots \otimes \vr_{d-1}$. The 
terms $\alpha_i \alpha_j$ correspond then to two-body correlations 
$\mean{\sigma_z^{(i-1)} \otimes \sigma_z^{(j-1)}}$ on the same state.  
On the other  hand, the right-hand side of Eq.~(\ref{eq-offdiagonal-appendix}) 
can be seen as a local hidden variable model, where the index ${\vec{a}}$ is the hidden 
variable occurring with probability $p_{\vec{a}}$, and the $a_i$ are 
the deterministic assignments for the measurement results of $\sigma_z$ 
on the different particles. For fully separable states, however, 
it is well known that all measurements can be explained by a local 
hidden variable model \cite{werner}. In fact, since one has a product
state, one can directly write down the model. For that, we define
\be
p_{\vec{a}} = \prod_{j=2}^d q_j(a_j),
\quad
\mbox{ where }
\quad
q_j(\pm 1) = \frac{1\pm\alpha_j}{2}.
\ee
This ensures that $q_j (+1)- q_j(-1) = \alpha_j$, moreover, due 
to the product structure, the two-body correlations are also 
reproduced.
\qed

\section*{Appendix B: Notes on the complexity of characterizing 
faithful entanglement}

In this section, we will show that the problem of characterizing
faithful entanglement is directly linked to the problem of so-called
unitary quadratic minimization (UQM). The latter problem was recently 
shown to be NP-hard, as the graph 3-coloring problem can be reduced 
to UQM \cite{lee2020}. 

Let us start by explaining the UQM problem. We start with a set of 
complex $n \times n$ matrices $A_1, \dots, A_k$ with a bounded 
Hilbert-Schmidt norm $\tr(A_j^\dagger A_j) \leq 1$ for all 
$j = 1, \dots, k.$ Then, the problem is to minimize the target 
function
\be
f(U) = \sum_{j=1}^k |\tr(A_j^\dagger U)|^2,
\label{eq-fdefinition}
\ee
where here and in the following $U$ always denotes an
$n \times n$  unitary matrix. More precisely, UQM is 
defined as a decision problem, where one is asked to decide
whether either 
\be
\min_U f(U) \leq \alpha \quad \mbox{ or } \quad \min_U f(U) \geq \alpha +\frac{1}{m} 
\ee
where $\alpha$ is a given real number and $m$ a given positive 
integer. As mentioned, this problem was shown to be NP-hard \cite{lee2020}. 

In order to understand the connection to the characterization 
of faithful entanglement, we consider an unnormalized maximally
entangled state $\ket{\Phi} = \sum_{i=1}^d \ket{ii}.$ Then, for 
an arbitrary matrix $A$ we define the state
\be
\ket{\Phi_A} = \eins \otimes A \ket{\Phi}.
\ee
Note that any pure state can be written in this form for
a suitable $A$. Furthermore, for an arbitrary unitary matrix $U$ we 
define
\be
\ket{\Phi_U} = \eins \otimes U \ket{\Phi}.
\ee
The state $\ket{\Phi_U}$ is maximally entangled and any maximally
entangled state can be written in this form. With these definitions,  
we have that
\be
\braket{\Phi_A}{\Phi_U} = \tr(A^\dagger U).
\ee
Applying this to Eq.~(\ref{eq-fdefinition}) one finds
that one can write
\be
f(U) = \bra{\Phi_U} X \ket{\Phi_U}
\label{eq-uqm1}
\ee
with
\be
X= \sum_{j=1}^k \ketbra{\Phi_{A_k}}.
\label{eq-uqm2}
\ee
Now the connection to faithful entanglement becomes apparent: As 
we have shown in Observation 1, in order to decide faithfulness 
one needs to maximize the overlap with maximally entangled states. 
Taking $X= (\eins -\varrho_{AB})/d$ (where the $d$ is due to the 
missing normalization of $\ket{\Phi_U}$) in Eqs.~(\ref{eq-uqm1}, 
\ref{eq-uqm2}), this leads to a minimization of $f(U)$ for a 
collection of $A_j.$ Of course, one may also start from a given 
collection of the $A_j$ and then compute the corresponding $\varrho_{AB}$.

In other words, if one is able to solve the maximization problem
$\max_U \bra{\Phi_U} Y \ket{\Phi_U}$ for general hermitean 
operators $Y$, then one can also solve the UQM problem described
above, which is known to be NP-hard. We note that the problem of
deciding faithfulness  is a little bit more restricted, as the
operator $Y$ is normalized and one is only interested in the question
whether the maximum is larger than $1/d$ or not. Still, the argument
suggests that deciding faithfulness is computationally hard. 

\section*{Appendix C: Additional numerical results}

\begin{table}
\begin{tabular}{|c||c|c|c|c|c|c|c|}
\hline
d & PPT & UFF [3(a)] &  UFF [3(b)] & FF [3(c)]\\
\hline
3 & 0.01\% &83.05\% & 94.55\% & 5.44\% \\ 
\hline
4 &0\% &99.93 \% &99.999\% & 0.001\% \\ 
\hline
5 &0\% & 100\% &100\% &0\% \\
\hline
6 &0\% & 100\% &100\% &0\% \\
\hline \hline
\end{tabular}
\caption{Fraction of states in the Hilbert-Schmidt distribution
that can be detected with the various criteria. First, we consider
the PPT states, which are always unfaithful. Then, we consider
the NPT states that are unfaithful (UFF) due to Observation 3(a).
Then, the NPT states that are unfaithful due to Observation 3(b). 
The fourth column is the fraction of states that can be detected
as faithful via Observation 3(c). }
\end{table}

As mentioned in the main text, we also considered random
states in the Hilbert Schmidt distribution in order to 
estimate which fraction is separable, entangled but unfaithful, 
or faithful. For two qubits we generated $10^6$ states and find 
$24.35\%$ of all states separable via the condition of the 
positivity of the partial transpose (PPT) \cite{hororeview}, and 
$21.14\%$ are entangled, but unfaithful, the remaining $54.51\%$ are faithful. As already explained in the main text, these values coincide up to statistical fluctuations with the values reported for the sufficient criterion SDP2 for unfaithfulness in Ref.~\cite{weilenmann20} (or Eq.~(\ref{eq-maxddim})). 
For higher dimensions, we also generated $10^6$ states randomly and 
the results are given in Table II.

\section*{Appendix D: Schmidt number witnesses}

Here, we discuss the notion of the Schmidt number
and Schmidt number witnesses. It turns out, however, that
Observation 1 cannot so easily be generalized to this 
case.

\subsection{The concept of the Schmidt number}

The concept of entanglement and separability can be generalized 
to the notion of the Schmidt number. For that, one defines states 
to be of Schmidt number $r$, if they can be written as a convex 
combination of pure states with Schmidt rank $r$, 
$
\varrho = \sum_k p_k \ketbra{\phi_k},
$
where all the states $\ket{\phi_k}$ have a Schmidt rank $r$. 
Clearly, the separable states are exactly the states with Schmidt 
number one. 

Then, Schmidt witnesses can be defined in analogy to entanglement 
witnesses: A Schmidt witness for Schmidt number $\ell$ is an observable 
$\SSS_\ell$ with positive expectation value on all states with Schmidt 
number $\ell$. So, observing a negative expectation value proves that the state has
at least Schmidt number $\ell+1$. Again, such witnesses can be based 
on the projector
\be
\SSS_\ell = \beta(\ell) \eins - \ketbra{\psi},
\label{eq-fidesch}
\ee
where the maximal squared overlap is now given by the sum of the 
$\ell$ biggest squared Schmidt coefficients,
\be
\beta(\ell) = \sum_{k=1}^\ell s_k^2.
\label{suml}
\ee
Again, the theory of Schmidt witnesses is well developed \cite{hulpke, sanpera, sarbicki}. The direct generalization of the RFW to the Schmidt witnesses 
is the set of Schmidt witnesses with $\ket{\psi}=\sum_{i=1}^d a_i \ket{ii}/\sqrt{d}$, 
where $a_i = \pm 1$. All these witnesses have $\beta(\ell) = r/d.$

\subsection{Fidelity-based Schmidt witnesses}

One may wonder whether a similar result as in Observation~1 
also holds for Schmidt witnesses. In the following, we show that 
a similar result holds for special cases, but not in general. 
Before going into details, we note a few general results on the 
order of witnesses for general convex sets.

We say a witness $\WW$ is weaker than a set of witnesses 
$\{\WW_k\}_k$, denoted as $\WW\prec\{\WW_k\}_k$, if for any $\varrho$ 
such that $\tr(\WW\varrho)<0$, there exists a $\WW_k$ such that 
$\tr(\WW_k\varrho)<0$, or equivalently, if 
$\tr(\WW_k\varrho)\ge0$ for all $k$, then $\tr(\WW\varrho)\ge0$. 
This relation can be analyzed by considering the following optimization 
problem:
\begin{align}
    \mbox{min: }&\tr(\WW\varrho)\label{eq:orderWit} \\
    \mbox{subject to: } & \tr(\WW_k\varrho)\ge 0  \mbox{ for all } k,\\
     &\varrho\ge 0. 
  \end{align}
More precisely, $\WW\prec\{\WW_k\}_k$ if and only if the solution of 
the  optimization in Eq.~\eqref{eq:orderWit} is zero. The dual problem of 
Eq.~\eqref{eq:orderWit} reads
\begin{align}
  \max_{x_k}\!:\;\;& 0 \\
  \mbox{subject to: } & \WW-\sum_kx_k\WW_k\ge 0,\label{eq:sumWit}\\
  & x_k\ge 0 \quad \text{for all~} k.
\end{align}
According to the general theory of convex optimization \cite{cvxopt}, 
weak duality gives that if there exists $x_k\ge 0$ such that Eq.~\eqref{eq:sumWit} 
holds, then $\WW\prec\{\WW_k\}_k$.  When the strong duality holds, e.g., in the case 
that $\{\WW_k\}_k$ is a finite set, then Eq.~\eqref{eq:sumWit} gives a necessary 
and sufficient condition. This can be viewed as a generalization of the result in Eq.~\eqref{eq-proof1}.

Now, we want to investigate whether any fidelity-based Schmidt witness 
$\SSS_\ell$ defined by Eqs.~\eqref{eq-fidesch} and \eqref{suml},
$
\SSS_\ell = \beta(\ell) \eins - \ketbra{\psi}
$ with 
$
\beta(\ell) = \sum_{k=1}^\ell s_k^2,
$
is always weaker than the set witnesses based on maximally entangled states, 
i.e, whether $\SSS_\ell\prec\{\SSS^\phi_\ell\}$. Here, $\{\SSS^\phi_\ell\}$ 
denotes the set of all Schmidt number $\ell$ witnesses based on the maximally 
entangled state:
\be
\frac{\ell}{d} \eins - U_A\otimes U_B\ketbra{\phi^+}
U_A^\dagger\otimes U_B^\dagger,
\ee
for all local unitary transformations $U_A$ and $U_B$. In order to see that 
there exists $\SSS_\ell$ such that $\SSS_\ell\not\prec\{\SSS^\phi_\ell\}$, we 
have the following observation.

{\bf Observation 4.} {\it Let $\ket{\psi}=\sum_{i=1}^ds_i\ket{ii}$ be 
a pure state with Schmidt number larger than $\ell$, i.e., $s_{\ell+1}>0$. 
Then it can be detected by $\{\SSS^\phi_\ell\}$ if and only if 
$\sum_{i=1}^ds_i>\sqrt{\ell}$.
}

This can be seen in the following manner: When maximizing
the overlap between two states over local unitaries, its best 
is to take them both in the same Schmidt basis \cite{guehnelutkenhaus}.
So, the maximization over all maximally entangled states 
$\ket{\phi}$ gives
\begin{equation}
  \max_{\ket{\phi}}|{\braket{\phi}{\psi}}|^2
  =\Big(\sum_{i=1}^d\frac{s_i}{\sqrt{d}}\Big)^2
  = \frac{1}{d}\big(\sum_{i=1}^d {s_i}\big)^2.
\end{equation}
The relation $\sum_{i=1}^ds_i>\sqrt{\ell}$ can always be satisfied 
when $\ell=1$. However, it can be violated when $\ell\ge 2$. On the 
other hand, $\ket{\psi}$ can always be detected by $\SSS_\ell$ defined 
by Eq.~\eqref{eq-fidesch}, as long as $s_{\ell+1}>0$. This implies that there
exists $\SSS_\ell$ such that $\SSS_\ell\not\prec\{\SSS^\phi_\ell\}$.
\qed

In general, we have the following necessary and sufficient condition for 
$\SSS_\ell\prec\{\SSS^\phi_\ell\}$:

{\bf Observation 5.} 
{\it The Schmidt witness $\SSS_\ell$ defined in Eq.~\eqref{eq-fidesch} 
satisfies that $\SSS_\ell\prec\{\SSS^\phi_\ell\}$ if and only if 
$s_1=s_2=\cdots = s_\ell$.
}

The sufficiency part follows similarly to Observation~1. To prove the 
necessity part, we assume that not all $s_1,s_2,\dots,s_\ell$ are equal
and construct a state that is detected by $\SSS_\ell$, but not by any
$\SSS_\ell^\phi$. First, since not all $s_1,s_2,\dots,s_\ell$ are equal,
we have $(\sum_{i=1}^\ell s_i)/\sqrt{\sum_{i=1}^\ell s_i^2} <\sqrt{\ell}$.
So, there exists ${0<\varepsilon\le s_{\ell+1}}$, such that
\begin{equation}
  \frac{1}{\sqrt{\sum_{i=1}^\ell s_i^2+\varepsilon^2}}\left(
  \sum_{i=1}^\ell s_i+\varepsilon\right)\le\sqrt{\ell}.
  \label{eq:ineqx}
\end{equation}
We define the state
$
  \ket{x}=\sum_{i=1}^{\ell+1}x_i\ket{ii},
$
with the coefficients
\begin{equation}
  \begin{aligned}
    &x_i=\frac{s_i}{\sqrt{\sum_{i=1}^\ell s_i^2+\varepsilon^2}}~\text{~for~}i=1,2,\dots,\ell,\\
    &x_{\ell+1}=\frac{\varepsilon}{\sqrt{\sum_{i=1}^\ell s_i^2+\varepsilon^2}},
  \end{aligned}
\end{equation}
Then $\ket{x}$ cannot be detected by $\{\SSS^\phi_\ell\}$ due to
Eq.~\eqref{eq:ineqx} and Observation~4. On the other hand, by 
taking advantage of the relation that $0<\varepsilon\le s_{\ell+1}$, 
we have
\begin{equation}
  |\braket{\psi}{x}|^2=\left(\sum_{i=1}^{\ell+1}s_ix_i\right)^2\ge
  \sum_{i=1}^\ell s_i^2+\varepsilon^2=\beta(\ell)+\varepsilon^2.
\end{equation}
This means that $\ket{x}$ can be detected by $\SSS_\ell$ but not
by witnesses in the set $\{\SSS^\phi_\ell\}$.
\qed


\end{document}